\documentclass[lettersize,journal]{IEEEtran}
\usepackage{amsmath,amsfonts}
\usepackage{algorithmic}
\usepackage{algorithm}
\usepackage{array}
\usepackage[caption=false,font=normalsize,labelfont=sf,textfont=sf]{subfig}
\usepackage{textcomp}
\usepackage{stfloats}
\usepackage{url}
\usepackage{verbatim}
\usepackage{graphicx}
\usepackage{cite}
\usepackage{times}
\usepackage{soul}
\usepackage{url}
\usepackage[hidelinks]{hyperref}
\usepackage[utf8]{inputenc}
\usepackage{graphicx}
\usepackage{amsmath}
\usepackage{amsthm}
\usepackage{booktabs}
\usepackage{algorithm}
\usepackage{algorithmic}
\usepackage[switch]{lineno}
\usepackage{textcomp}
\usepackage{xcolor}
\usepackage{multirow}
\usepackage{lipsum}
\usepackage{makecell} 
\usepackage{adjustbox}
\definecolor{Revision}{RGB}{0, 60, 255}
\definecolor{Response}{RGB}{0, 0, 139}
\definecolor{Letter}{RGB}{0, 60, 255}
\definecolor{summary}{RGB}{0, 60, 255}

\begin{document}

\title{STPNet: Scale-aware Text Prompt Network for Medical Image Segmentation}

\author{Dandan Shan, Zihan Li, 
        Yunxiang Li, Qingde Li, Jie Tian,~\IEEEmembership{Fellow,~IEEE}, Qingqi Hong,~\IEEEmembership{Member,~IEEE}
\thanks{Dandan Shan and Qingqi Hong are with Xiamen University, Xiamen 361005, China.}
\thanks{Zihan Li is with the Department of Bioengineering, University of Washington, Seattle, WA 98195, USA.}
\thanks{Yunxiang Li is with the Department of Radiation Oncology, UT Southwestern Medical Center, Dallas, TX 75235, USA.}
\thanks{Qingde Li is with the Department of Computer Science, University of Hull, Hull, HU6 7RX, UK.}
\thanks{Jie Tian is with the Institute of Automation, Chinese Academy of Sciences, Beijing, 100190, China.}
\thanks{Dandan Shan and Zihan Li have equal contributions to this work.}}

\markboth{IEEE TRANSACTIONS ON IMAGE PROCESSING,~Vol.~14, No.~8, August~2021}%
{Shell \MakeLowercase{\textit{et al.}}: A Sample Article Using IEEEtran.cls for IEEE Journals}


\maketitle

\begin{abstract}

{Accurate segmentation of lesions plays a critical role in medical image analysis and diagnosis. Traditional segmentation approaches that rely solely on visual features often struggle with the inherent uncertainty in lesion distribution and size. To address these issues, we propose STPNet, a Scale-aware Text Prompt Network that leverages vision-language modeling to enhance medical image segmentation. Our approach utilizes multi-scale textual descriptions to guide lesion localization and employs retrieval-segmentation joint learning to bridge the semantic gap between visual and linguistic modalities. Crucially, STPNet retrieves relevant textual information from a specialized medical text repository during training, eliminating the need for text input during inference while retaining the benefits of cross-modal learning. We evaluate STPNet on three datasets: COVID-Xray, COVID-CT, and Kvasir-SEG. Experimental results show that our vision-language approach outperforms state-of-the-art segmentation methods, demonstrating the effectiveness of incorporating textual semantic knowledge into medical image analysis.}
The code has been made publicly on \href{https://github.com/HUANGLIZI/STPNet}{https://github.com/HUANGLIZI/STPNet}.
\end{abstract}


\section{Introduction}

\label{sec:intro}

\IEEEPARstart{I}{n} the rapidly evolving field of healthcare, accurate segmentation of lesions remains a persistent challenge with profound implications for medical imaging. Manual annotation of medical images is not only time-consuming but also costly, as it requires expert clinicians to perform precise pixel-level labeling. To address this challenge, deep learning-based methods have gained prominence, leveraging advanced neural networks to achieve high-quality image segmentation and analysis with minimal reliance on extensive manual annotations. This significantly enhances efficiency and scalability. However, the inherent characteristics of medical images---such as low contrast, ambiguous boundaries, and limited dataset availability---pose significant challenges in training deep learning models for accurate lesion segmentation.

\begin{figure}
\centering 
\includegraphics[width= \linewidth]{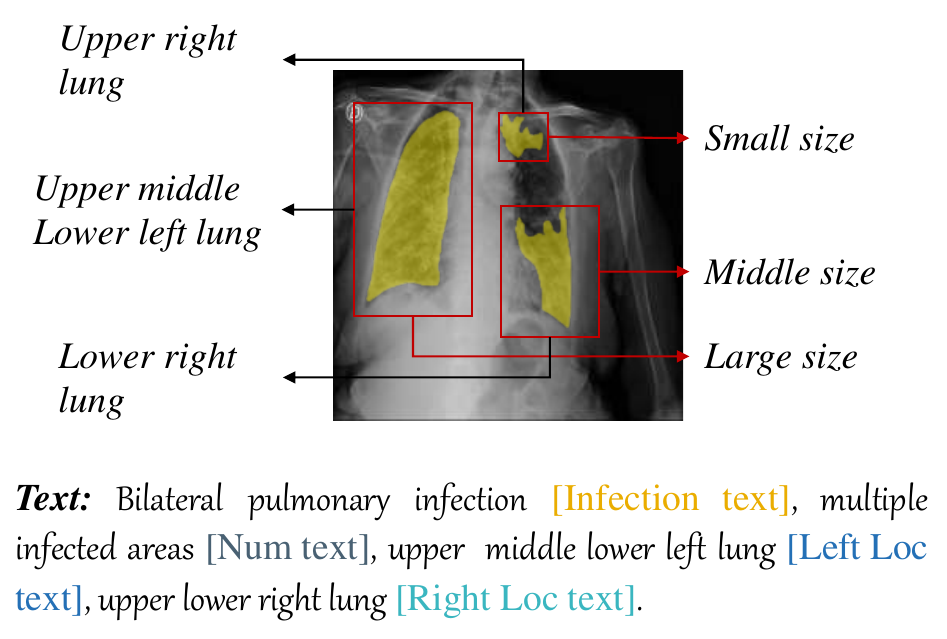}
\caption{Distribution of COVID-19. Lesions may be distributed in different locations with different sizes. The corresponding text labels for the images can be categorized into four types: Infection text, Num text, Left Loc text and Right Loc text. STPNet considers different locations and different scales jointly.}
\label{intro}
\vspace{-4mm}
\end{figure}

Additionally, the uncertainty associated with lesion presence and their potential occurrence in any anatomical region and size, further complicates the segmentation task. Current studies often fail to address these challenges comprehensively, as they typically focus on isolated aspects. Many researchers \cite{fan2020inf,zhao2021scoat,huang2021graph,li2022tfcns,xu2022mrdff} have primarily concentrated on distinguishing infected areas from healthy tissue but struggle with the effective segmentation of small lesions. Although some efforts have been made to address lesion boundaries \cite{huang2021graph}, segmentation performance remains limited due to factors such as image quality and the scarcity of annotated data. To improve segmentation accuracy, numerous approaches have adopted multi-scale learning techniques \cite{yan2021covid,zheng2020msd,pei2021mps} by utilizing convolutional kernels of different sizes to handle features at various scales. However, accurately segmenting lesion regions across different scales remains challenging due to irregular morphologies and the presence of hard-to-detect noise. In particular, smaller lesions are often misclassified as background, while larger lesions may also be incorrectly identified, further limiting segmentation effectiveness.


{Recent advances in vision-language modeling \cite{10172039, tomar2022tganet, huemann2024contextual, zhong2023ariadne} have opened new avenues for medical image analysis by leveraging the complementary nature of visual and textual information. Several pioneering works have explored this direction. For example, LViT \cite{10172039} utilizes BERT to encode medical reports and integrates this textual knowledge into Vision Transformer models to enhance segmentation performance. Similarly, TGANet \cite{tomar2022tganet} incorporates textual descriptions of polyp size attributes to improve the model's ability to generalize across polyps of varying dimensions. ConTEXTual Net \cite{huemann2024contextual} extracts linguistic features from radiology reports using pre-trained language models and fuses them with visual representations via cross-attention mechanisms, thereby enhancing pneumothorax segmentation accuracy in chest X-rays. Despite these advancements, current vision-language approaches for medical image segmentation face a \textbf{significant limitation}, i.e.,they require paired textual data during both training and inference phases. This constraint limits their applicability in real-world clinical scenarios, where such textual information may not always be available. \textbf{To overcome this constraint}, we propose a novel framework that leverages the power of vision-language modeling during training while maintaining inference efficiency \textbf{without textual input}. As shown in Fig. \ref{intro}, our approach involves developing a specialized medical text repository that encodes critical lesion characteristics: infection distribution (unilateral or bilateral), lesion quantity, and specific spatial localization information for both left and right anatomical regions. Through contrastive learning between image and text representations, we optimize cross-modal similarity, bringing semantically related image-text pairs closer in the embedding space while pushing unrelated pairs apart. This design allows our model to implicitly incorporate textual semantic knowledge during training while eliminating the need for explicit textual input during inference, thus significantly enhancing model adaptability and clinical practicality.}

To address the challenge of lesions appearing ``anywhere", our method focuses not only on the images themselves but also delves into leveraging information from medical reports. We believe that clinically relevant information extracted from these reports can provide valuable insights, aiding in lesion localization and thus enhancing segmentation performance. Additionally, to tackle the ``anysize" issue, we adopt a strategy combining complementary scale-semantic textual and visual features. To further enhance multi-scale representation, we introduce a Spatial Scale-Aware Module (SSM) designed to effectively capture spatial and multi-scale features. By seamlessly integrating textual and visual data, we aim to leverage the complementarity of these modalities to improve both the accuracy and robustness of lesion segmentation.

In this paper, we introduce STPNet, {a novel scale-aware textual prompt network} for medical image segmentation. Our approach leverages contrastive learning to extract rich information from medical reports and image data, eliminating the need for additional textual input during inference. Furthermore, by integrating complementary textual cues and visual features at the scale-semantic level within the segmentation network, our method addresses the challenge of accurately segmenting lesions of different sizes and irregular shapes. Additionally, our framework considers the spatial distribution of lesions, providing valuable insights to further enhance segmentation performance.

Our main contributions are summarized as follows:

\begin{itemize}
    \item We propose an innovative framework for joint medical text retrieval and image segmentation, named STPNet, which effectively addresses multi-modality and multi-scale challenges, achieving efficient lesion segmentation \textbf{without requiring textual input during inference}.
    \item We construct multi-scale textual features to compensate for the loss of image feature information caused by the gradual reduction in scale in U-shaped networks. Additionally, we propose Multi-Scale Textual Guidance Blocks (MTBlocks) and UTrans learning methods, enabling the network to jointly perceive and utilize both image and textual information.
    \item We evaluate STPNet on two pneumonia datasets (COVID-X-ray and COVID-CT) and one polyp dataset (Kvasir-SEG). Experimental results demonstrate that our STPNet significantly outperforms current state-of-the-art methods in segmentation tasks.
\end{itemize}

\begin{figure*}[!ht]
\centering 
\includegraphics[width=\linewidth]{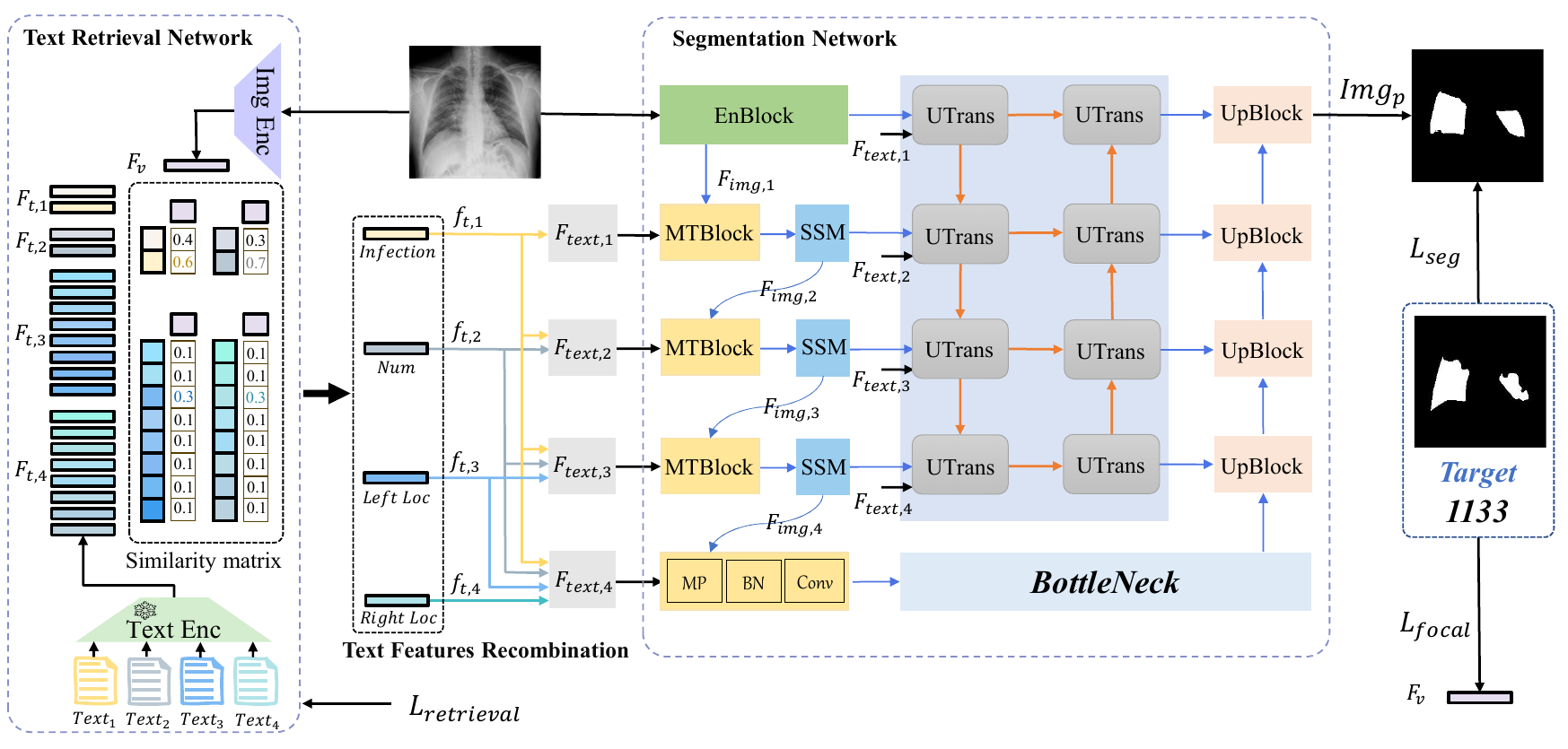}
\caption{Overview of our proposed STPNet. The framework consists of a Text Retrieval Network and a Segmentation Network. The Text Retrieval Network retrieves relevant text features by computing cosine similarity with image features. These retrieval text features are then recombined and fed into the Segmentation Network. The Segmentation Network, a hybrid CNN-Transformer architecture, employs MTBlock and UTrans for local and global encoding of text and image features. We utilize the Spatial Scale-aware Modules (SSM) to learn multi-scale and spatial features. Finally, the mixed-learning objectives optimize both the image segmentation and retrieval processes.}
\label{STPNet}
\vspace{-4mm}
\end{figure*}

\section{Related Works}
\label{sec:relatedworks}
\textbf{Medical Image Segmentation:} Existing medical image segmentation methods are primarily based on two mainstream frameworks: UNet and Transformer. UNet\cite{ronneberger2015u} adopts an encoder-decoder structure, which helps capture features at different scales and improves segmentation accuracy. Many researchers have extended this structure and achieved satisfactory performance \cite{valanarasu2022unext,han2022convunext,zhou2018unet++,oktay2018attention,isensee2021nnu,cao2021swin,xiong2024sam2}.
{SAM2UNet\cite{xiong2024sam2} framework combines the Hiera encoder of SAM2\cite{ravi2024sam} with a classic U-Net decoder and incorporates adapters into the encoder to enable more efficient parameter fine-tuning.}
The other mainstream framework is the Transformer model, which has achieved remarkable success in natural language processing tasks and has also been widely applied in medical image segmentation \cite{you2022class,wu2023d,hatamizadeh2022unetr,chen2021transunet,cao2021swin}. The Transformer model utilizes self-attention mechanisms to model long-range dependencies between pixels in an image. By leveraging attention mechanisms, the Transformer can capture global contextual information and enhance segmentation accuracy. However, these methods are based solely on image inputs, which may limit the potential for performance improvement in segmentation. To address this limitation, we propose a multimodal fusion approach that integrates both image and textual information to enhance medical image segmentation performance. By combining information from both modalities, we can attain a more comprehensive understanding of the images, leading to improved segmentation accuracy and robustness.

{\bf Multi-scale Learning:}
Features at different scales encapsulate valuable information with varying levels of detail, making their inclusion crucial in segmentation tasks. To address this issue, Yan et al. \cite{yan2021covid} introduced the Progressive Atrous Spatial Pyramid Pooling (PASPP) method, which employs a combination of Atrous convolutions. By integrating Atrous convolutions, PASPP effectively captures multi-scale features, thereby enhancing segmentation performance. Similarly, in the pursuit of multi-scale information, MSD-Net \cite{zheng2020msd} employs pyramid convolution blocks. This strategy enables simultaneous processing of features at different scales, leading to a more comprehensive understanding of the image and improved segmentation accuracy. Bakkouri et al.\cite{bakkouri2023mlca2f} have proposed MLCA2F, a framework that leverages multi-scale contextual information fusion to effectively capture critical lesion boundaries. Liu et al. \cite{LIU2023107493} proposed MESTrans that learns multi-scale information by incorporating a multi-scale embedding block and a feature fusion module, enabling integration of global context and adaptively fusing features at different scales for improved medical image segmentation.
However, these methods primarily focus on leveraging image features while they have not attempted to extract textual features at different scales from the more valuable radiology reports.

{\bf Text-guided Learning:}
In medical image segmentation, particularly for lesion segmentation, integrating text-guided learning offers a novel solution to address the limitations of traditional image-only methods. VTL \cite{ding2022vlt} and LAVT \cite{yang2022lavt} focus on natural image segmentation, {while UniLSeg \cite{liu2024universal} combines image and text to improve adaptability across tasks, performing well with multi-task data and automatic annotation. Recently, CMIRNet\cite{xu2024cmirnet} introduces a cross-modal interaction network that uses language queries for precise object segmentation.} Generally, these methods heavily rely on language encoders to explicitly align text and image data. However, the complexity and ambiguity of medical images present significant challenges for such strict alignment strategies. Consequently, tailored approaches have been introduced to address the unique requirements of medical imaging, incorporating text information at various stages of the encoder-decoder structure. For instance, LViT \cite{10172039} integrates text during the encoder stage to enhance performance, while C2FVL \cite{shan2023coarse} and VLSM \cite{poudel2023exploring} introduce text features at the skip connections. ConTEXTual Net \cite{huemann2024contextual} and TGANet\cite{tomar2022tganet} incorporate text in the decoder. Additionally, models like CLIP\cite{radford2021learning}, GLoRIA\cite{huang2021gloria}, and ConVIRT\cite{zhang2020contrastive}, pre-trained on large-scale image-text pairs, enable image encoders to learn textual information, improving performance in downstream segmentation tasks.
However, the above methods rely on text input, which will limit the universality of the model during the inference stage and may cause a loss of model performance due to the diversity of text input and inaccurate text input.

\section{Method}
\label{sec:method}


\subsection{Overview of STPNet}

{As illustrated in Fig. \ref{STPNet}, the STPNet framework seamlessly integrates vision-language modeling for medical image segmentation through a dual-network architecture consisting of a text retrieval network and an image segmentation network. The text retrieval network employs contrastive learning to measure the similarity between image and text features, retrieving the most relevant textual descriptions via an image encoder paired with a frozen pre-trained text encoder. This alignment enables the retrieval of pertinent textual information that encodes lesion characteristics (distribution patterns, quantity, and spatial localization) without requiring explicit text input during inference. The segmentation network then leverages this retrieved textual knowledge through a hierarchical integration mechanism, following a coarse-to-fine feature fusion approach that combines convolutional neural network blocks (MTBlock, SSM, UpBlock) with transformer modules (UTrans). After initial feature extraction via EnBlock, the MTBlock receives both image features and retrieved textual features, performing early cross-modal fusion that enables the model to leverage linguistic semantic cues to guide the visual segmentation process. Unlike conventional vision-language models that employ separate encoders followed by cross-modal attention, STPNet integrates textual information directly into the image encoding process, addressing the scale variability challenge in lesion segmentation by utilizing textual descriptions that implicitly encode scale information at multiple levels of the feature hierarchy, thereby achieving robust performance across lesions of varying sizes and distributions.}

\begin{figure*}
\centering 
\includegraphics[width=0.78\linewidth]{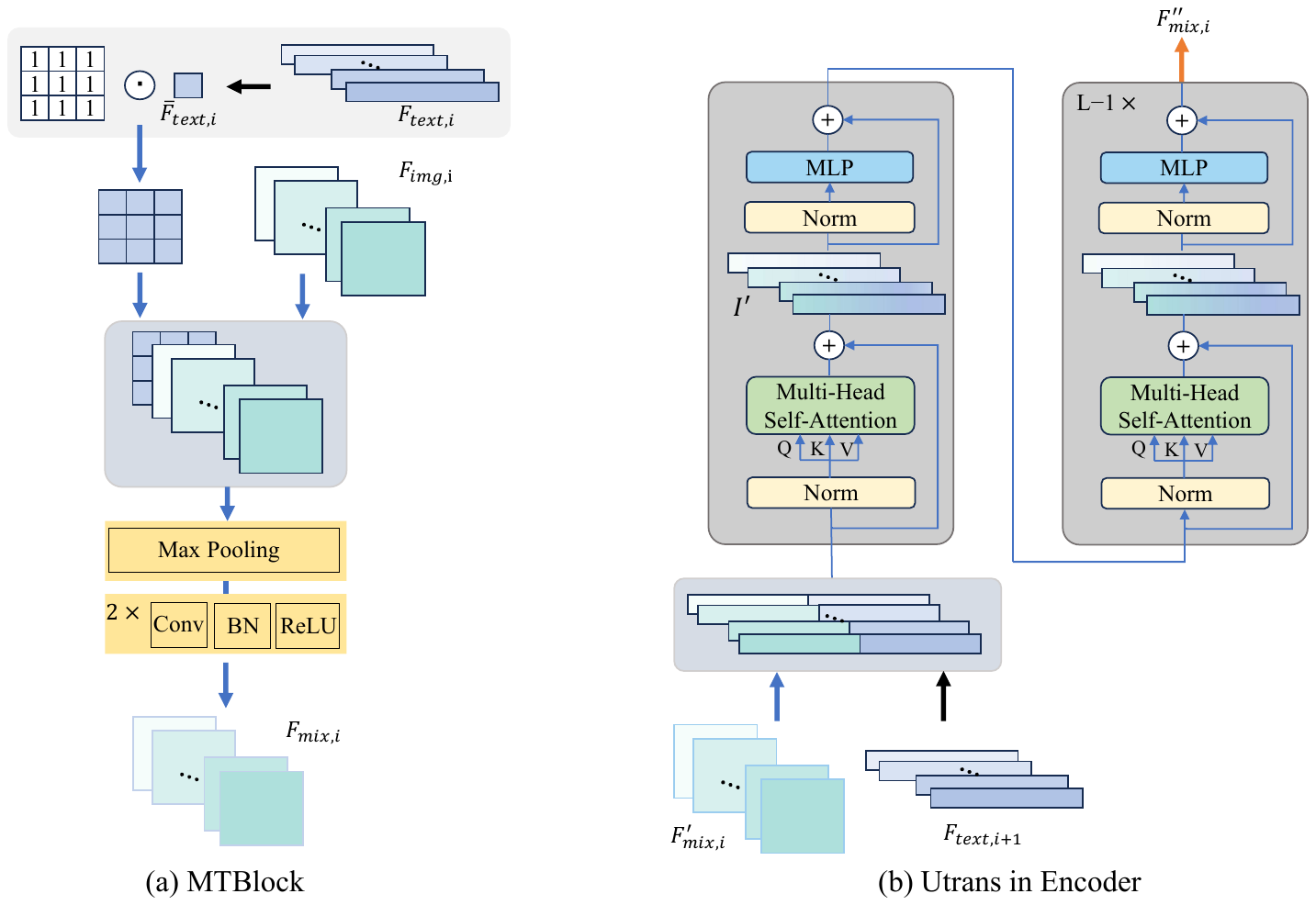}
\caption{Illustration of MTBlock and UTrans Encoder. (a) In the MTBlock integration process, the average text features $\bar F_{text,i}$ are concatenated with image features $F_{text,i}$ after feature map formation. This combined representation undergoes convolution to produce fused visual-textual features  $F_{mix,i}$. (b) In the UTrans Encoder, image features $F_{mix,i}$ processed by SSM are merged with text features $F_{text,i+1}$ using a multi-head self-attention mechanism. And, only the image-related features are extracted and inputted into the subsequent transformer. }
\label{ITfusion}

\end{figure*}
\subsection{Text Retrieval Network}
\textbf{Image Encoder.}
 Considering the robust image feature encoding capabilities of ResNet-101 \cite{he2016deep}, we utilize the pre-trained ResNet-101 network as the backbone to construct the image encoder $Enc_{v}$ during the training process. 
Features extracted from layer 4 undergo an upsampling process and are then concatenated with features from layer 3, creating a multi-scale feature representation. These multi-scale features are subsequently processed through MLP for further mapping. Ultimately, an image feature representation is obtained via a MaxPooling (MP) operation, thereby enhancing the expressiveness of the image features.
\begin{eqnarray}
& {v}_{h} = {Enc}_{v}(Img) \\
& {F}_{v}=MP(MLP([Upsample({v}_{h});{v}_{l}])) 
\end{eqnarray}
Where, ${v}_{l}$ represents the features extracted from Layer 3 of ResNet-101, ${v}_{h}$ denotes the features from Layer 4.

\textbf{Text Encoder.}
For an image, we have four types of text database, including ``Infection text" $Text_{1}$ (describing  whether the lesion is bilateral/unilateral), ``Num text" $Text_{2}$ (describing  the number of lesions), ``Left Loc text" $Text_{3}$ and ``Right Loc text" $Text_{4}$ (respectively describing  the distribution of the lesion's position on the left and right sides). To ensure performance, we have chosen the frozen BioClinicalBERT \cite{alsentzer2019publicly} model as our text encoder $Enc_{t}$. This model has been pre-trained on large-scale clinical text data and possesses strong capabilities in learning text representations. Different types of text descriptions are fed into the text encoder to obtain corresponding text feature representations:
\begin{eqnarray}
& {F}_{t,i} = {Enc}_{t}(Text_{i}), i\in \left \{ 1,2,3,4\right \}
\end{eqnarray}


\textbf{Text Retrieval.}
We bridge $F_{t,i}$ and $F_v$ by computing the cosine similarity matrix. Specifically, cosine similarity evaluates their similarity by measuring the cosine of the angle between two vectors. The closer the value is to 1, the more similar the vectors; the closer to 0, the less similar they are. The similarity calculation is represented as follows:
\begin{eqnarray}
& s_{ij} = \frac{(F_v \cdot F_{t,i,j})}{\|F_v\| \cdot \|F_{t,i,j}\|} \\
& j^{*} = \text{argmax}_{j} \left( s_{ij} \right) \\
& f_{t,i} = F_{t,i,j^{*}}
\end{eqnarray}
The $s_{ij}$ scores are used to rank all sentences $F_{t,i,j}$ within each $F_{t,i}$ and the top-ranked candidate sentence $f_{t,i}$, where $f_{t,i} \in F_{t,i}, i \in \left \{1,2,3,4 \right \}$  is returned as the retrieved text feature. Through this approach, we can effectively find the text feature that is most compatible with the queried image.


\subsection{Text Features Recombination}

{Due to the characteristics of the U-shaped network structure, increasing network depth leads to a progressive reduction in the image scale, resulting in the loss of fine-grained details. To mitigate this issue, we introduce multi-scale text features, which refers to a semantic hierarchy of textual information structured at different levels of granularity and scope, explicitly designed to parallel the spatial feature hierarchy in the visual pipeline. Each text feature encapsulates information at a distinct semantic scale, ensuring a comprehensive representation of lesion characteristics. Our approach utilizes four distinct types of text features that capture different semantic aspects without spatial repetition. The ``Infection text" feature provides macroscopic, global information about infection distribution patterns (unilateral/bilateral). The ``Num text" feature offers more specific, quantitative descriptions of lesion prevalence. Meanwhile, the ``Left Loc text" and ``Right Loc text" features contain localized positional information for infected regions. Together, these text features form a structured progression from global to increasingly localized descriptions, enriching the segmentation process with hierarchical semantic cues.}

Subsequently, these diverse text features undergo a recombination process, transitioning from coarse to fine granularity. Sequentially integrating the ``Infect text," ``Num text," ``Left Loc text," and ``Right Loc text," they culminate in the formation of increasingly rich semantic text features, denoted as $F_{text,i}$.
This rich semantic information is then seamlessly integrated into various layers of the MTBlock and UTrans components within the U-shaped network. By fostering a symbiotic relationship between image and text features, we aim to bolster the network's overall performance and efficacy. we opt for mean value computation to recombine these text features because it helps balance the information weight among different types of features, allowing each feature to contribute to the final result and forming more representative text features. The specific recombination process is illustrated by the following equations:
\begin{eqnarray}
F_{\text{text},i} = \frac{1}{i} \sum_{j=1}^{i} f_{t,j}
\end{eqnarray}

\begin{figure}
\centering 
\includegraphics[width=1\linewidth]{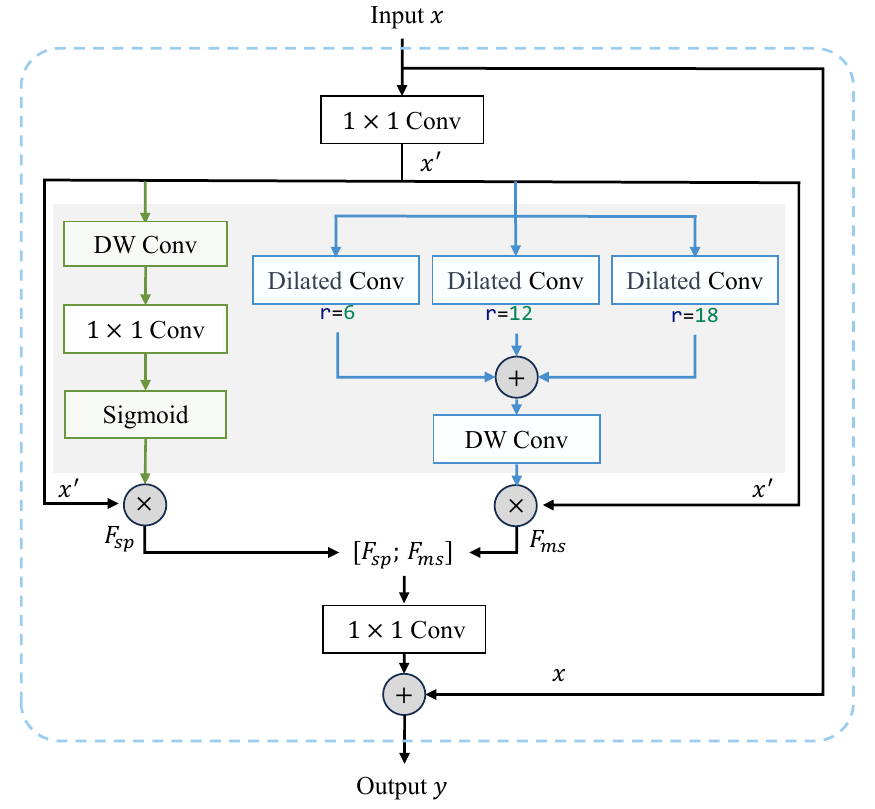}
\caption{The structure of the SSM (Spatial Scale-aware Module) involves learning multi-scale features through convolutions with different dilation rates and integrating spatial features using 1x1 convolutions.}
\label{SSA}

\end{figure}
\vspace{-5mm}
\subsection{Segmentation Network}
\textbf{Multi-scale Text-guided Block (MTBlock).}
We designed a new Multi-scale Text-guided block (MTBlock) to enable interaction between image features and text features.  First, $F_{img,i}$ is encoded by EnBlock, as shown in Equation \ref{3.2-1}. 
The specific image-text fusion process is shown in Fig. \ref{ITfusion}(a). For the text feature $F_{text,i} $, we compute the global average to obtain a feature that captures the overall semantic context of the text. This averaged text feature $\bar F_{text,i} $ is then multiplied by an all-ones matrix to form a feature map with the same height and width as the image feature. These two feature maps, $F_{img,i}$ and $\bar F_{text,i}$, are concatenated to combine visual and textual information. Subsequently, the combined features undergo Max Pooling, Batch Normalization (BN), and Convolution operations. ReLU is used for activation to introduce non-linearity, resulting in the final output $F_{mix,i}$ containing the semantic information of both image and text.

\vspace{-1mm}
\begin{eqnarray}
& F_{img,i} = EnBlock(Img)\label{3.2-1}\\
&\bar{F}_{\text{text},i} = \frac{1}{L \times D} \sum_{j=1}^{L} \sum_{k=1}^{D} F_{\text{text},i,j,k}\\
& F^{}_{mix,i} = Conv(BN(MP([F_{img,i}; \mathbf{1} \odot \bar{F}_{\text{text},i}])))\label{3.2-4}
\end{eqnarray}

Where, $L$ and $D$ represent the length of the text and the length of the features, respectively.

\textbf{Spatial Scale-aware Module (SSM).}
To enhance the model's perception of multi-scale spatial information in the fused image-text features, we designed the Spatial Scale-aware Module (SSM), as shown in Fig. \ref{SSA}. Specifically, the input feature $x$ first undergoes a $1\times1$ convolution to obtain the feature $x^{'}$, followed by two key steps: spatial feature extraction and multi-scale feature extraction. In terms of spatial feature extraction, the module employs deep-wise convolution and $1\times1$ convolution to weight $x^{'}$, and activates it using the Sigmoid function. This enables the model to focus on important regions while suppressing irrelevant parts, thereby enhancing the model's perception of spatial information in images and text. Meanwhile, in the multi-scale feature extraction aspect, the SSM module utilizes dilated convolutions with different dilation rates to enable the model to understand the fused multi-scale image-text information comprehensively. These dilated convolution operations help the model capture features at different scales, thereby improving its understanding of multi-scale data. Finally, the features obtained from spatial feature extraction and multi-scale feature extraction are concatenated and combined with the input feature $x$ through residual connections to obtain the final output feature. The detailed process is illustrated in Eqn. \ref{eqn11}-\ref{eqn12}.
\begin{eqnarray}
& {x}^{'} = {Conv}_{1}(x) \label{eqn11}\\
& {F}_{sp} = {x}^{'} \bigodot Sigmoid({Conv}_{1}({Conv}_{dw}( {x}^{'}))) \\
& {F}_{msa} = \text{{Conv}}_{dt1}({x}^{'}) + \text{{Conv}}_{dt2}({x}^{'}) + \text{{Conv}}_{dt3}({x}^{'})\\
& {F}_{ms} = {x}^{'} \bigodot {Conv}_{dw} ( {F}_{msa} )  \\
& y = x + {Conv}_{1}([{F}_{sp}; {F}_{ms}]) \label{eqn12}
\end{eqnarray} 

Where, ${Conv}_{1}$ represents $1\times1$ convolution, ${Conv}_{dw}$ denotes depth-wise convolution, and ${Conv}_{dt1}$, ${Conv}_{dt2}$, ${Conv}_{dt3}$ represent the dilation convolutions with dilation rates of 6, 12, and 18 respectively.

\textbf{Structure of UTrans.}
As shown in Fig. \ref{ITfusion} (b), the proposed UTrans follows the structure of the Vision Transformer, with modifications made only to the input and output sections of the Multi-Head Self-Attention mechanism in the encoding part. Inspired by \cite{hu2023animate}, in this architecture, image features and text features are concatenated, normalized, and fed into the self-attention mechanism. However, only the image features are retained after the self-attention mechanism. This design ensures that the image features also incorporate information from the text, enhancing the model's ability to learn from both modalities simultaneously and consistently. The specific process is as follows:
\begin{eqnarray}
& F_{mix,i}^{'} =  \left[ {i}_1, {i}_2, \ldots, {i}_N \right] \\
& F_{text,i+1} = \left[  {t}_1, {t}_2, \ldots, {t}_M \right] \\
& X = Norm(\left[ F_{mix,i}^{'}; F_{text,i+1}\right]) \\
& {Q} = {X} {W}_Q, \quad {K} = {X} {W}_K, \quad {V} = {X} {W}_V \\
& {[I',T'] = {Attention}({Q}, {K}, {V}) = \text{softmax}\left( \frac{{Q} {K}^T}{\sqrt{d_k}} \right) {V}} \\
&{I}' = [{i}'_1, {i}'_2, \ldots, {i}'_N]\\
&{{T}' = [{t}'_1, {t}'_2, \ldots, {t}'_M]}
\end{eqnarray}

Where $F_{mix,i}^{'}$ and $F_{text,i+1}$ represent the sequence of image features after passing through SSM and text features, respectively. X represents the concatenated sequence of image and text features, 
{$I'$ and $T'$ represent the output sequences of image features and text features, respectively, after the self-attention mechanism.}
{Our approach differs from conventional cross-attention mechanisms that maintain separate modal representations. Instead, our concatenation-based strategy enables direct feature interaction across granularities while preserving crucial spatial information. This integration allows textual semantics to directly enhance visual features without disrupting spatial relationships, which is particularly beneficial for accurate lesion boundary delineation across varying scales and distributions in medical images.}

\subsection{Training Objectives}
As shown in Fig. \ref{STPNet}, our training objective consists of $\operatorname{L}_{seg}$, $\operatorname{L}_{retrieval}$ and $\operatorname{L}_{focal}$. The segmentation loss $\operatorname{L}_{seg}$ consists of the Dice Loss and Cross Entropy Loss as shown in Eqn. \ref{3.4-1}, where $p_{i}$ represents the prediction result of the i-th pixel, and $y_{i}$ represents the ground truth (GT) value of the corresponding pixel. 
\begin{eqnarray}
&{\operatorname{L}_{seg}}=1-\frac{1}{N}\sum_{i=1}^{N}({\frac{2\left|p_{i}\cap y_{i}\right|}{\left(\left|p_{i}\right|+\left|y_{i}\right|\right)}}+{}y_{i}\log{\left(p_{i}\right)})\label{3.4-1}
\end{eqnarray}
During the retrieval process, we utilize the retrieval loss function $L_{retrieval}$ to minimize the distance between corresponding image features and text features.
\begin{eqnarray}
L_{retrieval} = \sum_{i=1}^{4} -\log \frac{\exp(F_{t,i} \cdot F_{v} / \tau)}{\sum_{n \in N(i)} \exp(F_{t,n} \cdot F_{v} / \tau)}
\end{eqnarray}
Where, $N(i)$ is the set of all samples for the i-th type of text database, including one positive sample and multiple negative samples. $\tau$ is the temperature parameter. $F_{t,n}$ represents the text feature used as either a negative or positive sample in the calculation. In addition, Focal loss is used to optimize the image feature $F_v$ based on the retrieved categories, and $\gamma$ is set as 2 in Eqn. \ref{3.4-f}.
\begin{eqnarray}
{\operatorname{L}_{focal}} = -(1 - softmax(F_{v}))^{\gamma}\log(softmax(F_{v})) \label{3.4-f}
\end{eqnarray}

The complete training objectives are shown in Eqn. \ref{3.4-3}, {where $\lambda_1 = \lambda_2 = \lambda_3 = 1$.}
\begin{eqnarray}
&{{\operatorname{L}_{mix}}=\lambda_1{\operatorname{L}_{seg}}+ \lambda_2{\operatorname{L}_{retrieval}}+ \lambda_3{\operatorname{L}_{focal}} \label{3.4-3}}
\end{eqnarray}

\section{Experimental Results and Discussion}
\label{sec:experimental}
\subsection{Datasets and Implementation Details}
\textbf{Datasets. COVID-CT} \cite{ma2020towards,morozov2020mosmeddata}:
This dataset combines 2729 slices from three publicly available datasets, each labeled by different radiologists with different labels. The different datasets were combined to map different colored masks to white uniformly.
\textbf{COVID-Xray} \cite{degerli2021covid}:
The dataset contains a total of 9258 chest X-ray images, the training set contains 7145, the test set contains 2113, and radiologists label the COVID-19 infection area of each image. 
\textbf{Kvasir-SEG} \cite{jha2020kvasir}: 
This dataset consists of 1000 high-quality gastrointestinal polyp images, openly accessible and paired with corresponding segmentation masks. These annotations have been meticulously crafted and validated by an experienced gastroenterologist.

\textbf{Implementation Details.} For the COVID-CT dataset, we randomly divide the train set, the validation set, and the test set according to the ratio of 8:1:1. For the COVID-Xray dataset, we adopt the official test set and divide the 20\% of data in the official train set as validation set.  For the Kvasir-SEG dataset, we adopted the official split of 880/120 for training and testing as the same setting of TGANet \cite{tomar2022tganet}. 

{To construct the text database, we use expert-annotated medical descriptions from each dataset. For the COVID-CT and COVID-Xray datasets, these descriptions summarize the distribution, number, and location of lesions. For example, a description might state: “Bilateral pulmonary infection, three infected areas, middle left lung and middle lower right lung."
We normalize all textual descriptions to maintain consistency. Specifically, we ensure that the lesion quantity is described as either “One infected area" or “Multiple infected areas" and standardize the “All" descriptions in Left Loc text and Right Loc text to “Upper middle lower."}

{To systematically structure this information, we categorize it into four key components: Infection text: [“Unilateral pulmonary infection",  “Bilateral pulmonary infection"]; 
Num text: [“One infected area", “Multiple infected areas"]; 
Left Loc text: [“No lesion in left lung", “Upper left lung", “Middle left lung", “Lower left lung", “Upper lower left lung",  “Upper middle left lung", “Middle lower left lung", “Upper middle lower left lung"];
Right Loc text: [“No lesion in right lung", “Upper right lung", “Middle right lung", “Lower right lung", “Upper lower right lung", “Upper middle right lung", “Middle lower right lung", “Upper middle lower right lung"].
For the Kvasir-SEG dataset, the term “lung" can simply be replaced with “polyp" to construct a corresponding text database.}

We implement our method based on the PyTorch framework and A100. We set the initial learning rate as 3e-4 and the batch size as 24. The number of early stop epochs we set is 100.

\begin{table}[!t]
\caption{Performance comparison between our proposed method (STPNet) and other state-of-the-art methods on COVID-CT dataset. Since our text information is retrieved by the model rather than directly added, the ``Text Input“ is displayed as ``$\times$”. }
\centering
\resizebox{0.4\textwidth}{!}{%
\begin{tabular}{cccccc}
\toprule
{\textbf{Method}} & {\textbf{Text Input}} & Dice(\%)     & IoU(\%)     \\ \midrule
UNet\cite{ronneberger2015u}                  & $\times$   & 62.96 & 50.10   \\ 
UNet++\cite{zhou2018unet++}                  & $\times$   & 71.75 & 58.39 \\ 
AttUNet\cite{oktay2018attention}             & $\times$   & 65.57 & 51.82 \\ 
nnUNet\cite{isensee2021nnu}                  & $\times$   & 72.59 & 60.36   \\ 
TransUNet\cite{chen2021transunet}            & $\times$   & 71.24 & 58.44  \\ 
Swin-UNet\cite{cao2021swin}                  & $\times$   & 63.29  & 50.19  \\

UCTransNet\cite{wang2022uctransnet}          & $\times$   & 66.74 & 52.56  \\ 
{SAM2UNet\cite{xiong2024sam2}}              & {$\times$}   & {72.91} & {60.21} \\
ConVIRT\cite{zhang2020contrastive}           & \checkmark & 72.06 & 59.73 \\
TGANet\cite{tomar2022tganet}                 & \checkmark & 70.29 & 57.38 \\ 
GLoRIA\cite{huang2021gloria}                 & \checkmark & 72.42 & 60.2  \\
CLIP\cite{radford2021learning}             & \checkmark & 71.97 & 59.64  \\
{ViLT\cite{28kim2021vilt}}                   & \checkmark & 72.36 & 60.15 \\
{LAVT\cite{29yang2021lavt}}                  & \checkmark & 73.29 & 60.41\\
{UniLSeg
\cite{liu2024universal}}   & {\checkmark} & {71.60} & {57.95} \\ 

{CMIRNet
\cite{xu2024cmirnet}}   & {\checkmark} & {73.69} & {60.44} \\

 \textbf{STPNet (ours)}                      & $\times$   &\textbf{76.18} & \textbf{63.41}  \\\bottomrule
\end{tabular}}
\label{COVID_CT}
\end{table}

\begin{table}[!t]
\caption{Performance comparison between our proposed method (STPNet) and other state-of-the-art methods on COVID-Xray dataset. Since our text information is retrieved by the model rather than directly added, the ``Text Input” is displayed as ``$\times$”. }
\centering
\resizebox{0.4\textwidth}{!}{%
\begin{tabular}{cccccc}
\toprule
{\textbf{Method}} & {\textbf{Text Input}} & Dice(\%)     & IoU(\%)      \\ \midrule
UNet\cite{ronneberger2015u}                  & $\times$   & 79.02 &	69.46   \\ 
UNet++\cite{zhou2018unet++}                  & $\times$   & 79.62 & 70.25   \\
AttUNet\cite{oktay2018attention}             & $\times$   & 79.31 &	70.04 
 \\ 
nnUNet\cite{isensee2021nnu}                  & $\times$   & 79.90 & 70.59 
 \\ 
TransUNet\cite{chen2021transunet}            & $\times$   & 78.63 & 69.13  
 \\ 
Swin-UNet\cite{cao2021swin}                  & $\times$   & 77.27 & 67.96  
 \\
UCTransNet\cite{wang2022uctransnet}          & $\times$   & 79.15 &	69.60   \\
{SAM2UNet\cite{xiong2024sam2}}              & {$\times$}   & {79.46} & {70.07}
 \\
ConVIRT\cite{zhang2020contrastive}           & \checkmark & 79.72 & 70.58 
 \\
TGANet\cite{tomar2022tganet}                 & \checkmark & 79.87 & 70.75 \\
GLoRIA\cite{huang2021gloria}                 & \checkmark & 79.94 & 70.68  
 \\
CLIP\cite{radford2021learning}             & \checkmark & 79.81 & 70.66  
 \\
{ViLT\cite{28kim2021vilt}}                   & \checkmark & 79.63 & 70.12 
 \\
{LAVT\cite{29yang2021lavt}}                  & \checkmark & 79.28 & 69.89
 \\
 {UniLSeg
\cite{liu2024universal}}   & {\checkmark} & {79.99} & {70.29} \\ 

{CMIRNet
\cite{xu2024cmirnet}}   & {\checkmark} & {79.36} & {69.98} \\

 \textbf{STPNet (ours)}                      & $\times$   &\textbf{80.63} & \textbf{71.42} \\\bottomrule
\end{tabular}}
\label{COVID_Xray}
\end{table}

\subsection{Comparison with State-of-the-Art Methods}
We compared our proposed STPNet method with other state-of-the-art methods, including those that do not utilize text input, as well as those that do, to thoroughly demonstrate its performance. As illustrated in Table \ref{COVID_CT}, \ref{COVID_Xray}, and \ref{KvasirSEG}, our method is superior to other methods in Dice score and IoU metric on all three datasets.
\begin{table}[!t]
\caption{Performance comparison between our proposed method (STPNet) and other state-of-the-art methods on Kvasir-SEG dataset. Since our text information is retrieved by the model rather than directly added, the ``Text Input” is displayed as ``$\times$”. }
\centering
\resizebox{0.4\textwidth}{!}{%
\begin{tabular}{cccccc}
\toprule
{\textbf{Method}} & {\textbf{Text Input}} & Dice(\%)     & IoU(\%) \\ 
\midrule
\vspace{0.1mm} 
UNet\cite{ronneberger2015u}         & $\times$   & 82.66 & 73.62 \\ 
\vspace{0.1mm} 
UNet++\cite{zhou2018unet++}         & $\times$   & 82.38 & 72.43  \\
\vspace{0.1mm} 
{AttUNet\cite{oktay2018attention} }                             & $\times$   & {85.72} & {78.12}\\
\vspace{0.1mm} 
nnUNet\cite{isensee2021nnu}         & $\times$   & 85.77 & 78.11 \\ 
\vspace{0.1mm} 
TransUNet\cite{chen2021transunet}            & $\times$   & 86.91 & 79.37  \\ 
\vspace{0.1mm} 
Swin-UNet\cite{cao2021swin}         & $\times$   & 85.90 & 77.56 \\
\vspace{0.1mm} 
UCTransNet\cite{wang2022uctransnet} & $\times$   & 86.73 & 79.09 \\
\vspace{0.1mm} 
{SAM2UNet\cite{xiong2024sam2}}       & {$\times$}   & {92.80} & {87.90} \\
\vspace{0.1mm} 
{ConVIRT\cite{zhang2020contrastive}}           & {\checkmark} & {78.50} & {68.59}  \\
TGANet\cite{tomar2022tganet}        & \checkmark & 89.82 & 83.30 \\
{GLoRIA\cite{huang2021gloria}}                 & {\checkmark} & {87.74} &  {80.58} \\
{CLIP\cite{radford2021learning}}             & {\checkmark} & {81.90} &  {71.38}   \\
{ViLT\cite{28kim2021vilt}}                   & {\checkmark} & {72.60} &  {59.38} \\
{LAVT\cite{29yang2021lavt}}                  & {\checkmark} & {89.98}  &  {84.37} \\
{UniLSeg
\cite{liu2024universal}}   & {\checkmark} & {90.95} & {84.57} \\

{CMIRNet
\cite{xu2024cmirnet}}   & {\checkmark} & {88.91} & {82.43}\\

\textbf{STPNet (ours)}              & $\times$   & \textbf{98.19} & \textbf{96.45} \\\bottomrule
\end{tabular}
}
\label{KvasirSEG}
\end{table}

\begin{figure}[!t]
\centering
  \includegraphics[width=\linewidth]{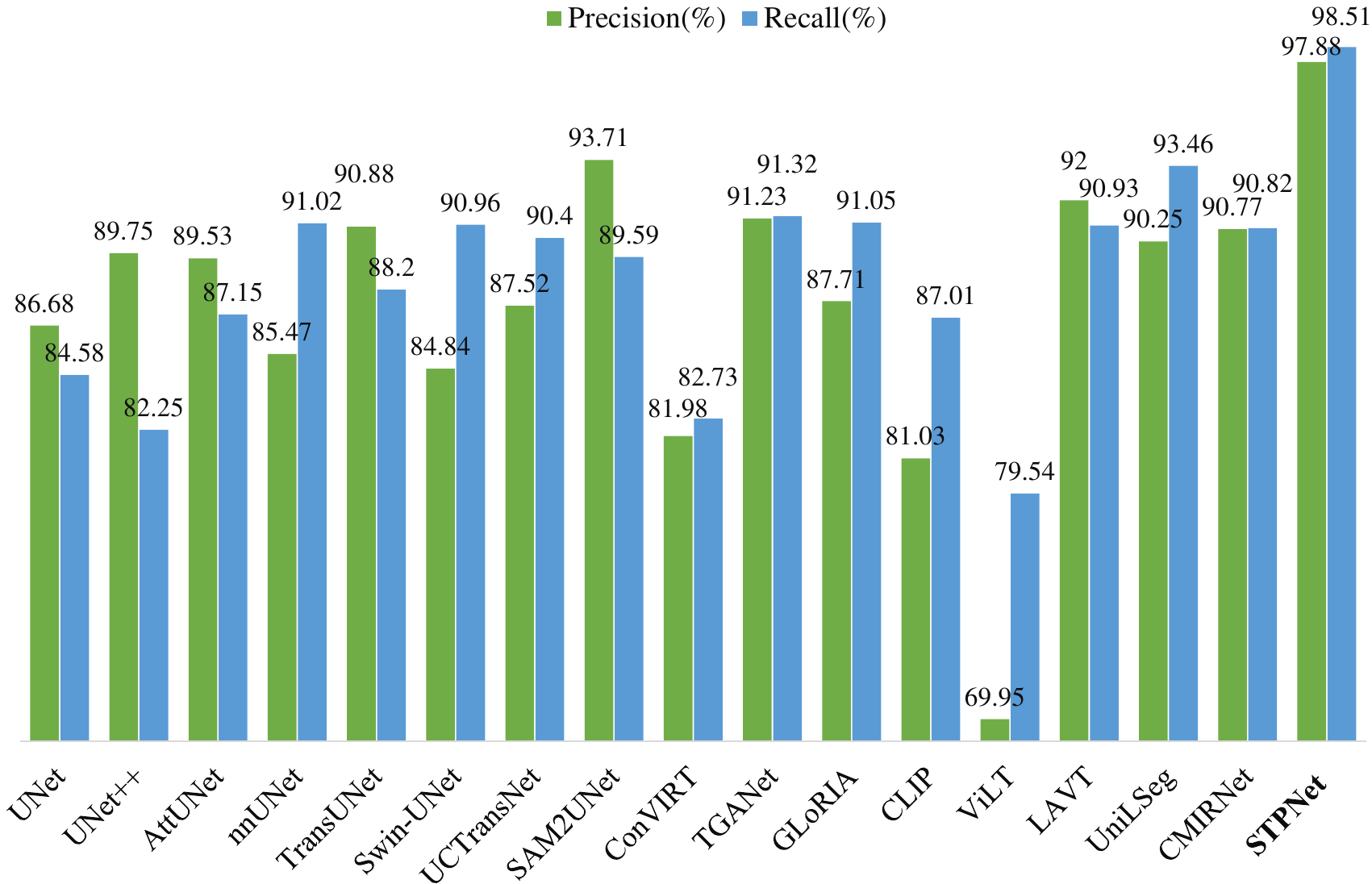}  \caption{{Performance comparison between our proposed method (STPNet) and other state-of-the-art methods on the Kvasir-SEG dataset using the Precision and Recall metrics.}}
  \label{hd95}
\end{figure}

{Specifically, in Table \ref{COVID_CT}, compared to SAM2UNet\cite{xiong2024sam2}, a benchmark framework based on foundation models, our method achieved an improvement of 3.27\% in Dice on the COVID-CT dataset. Additionally, compared to CMIRNet \cite{xu2024cmirnet}, a cross-modal interaction network, our model achieved an improvement of 2.49\% in Dice and 2.97\% in IoU, without directly incorporating text as input. In Table \ref{COVID_Xray}, our model outperformed the second best method, i.e.,UniLSeg \cite{liu2024universal}, by an improvement of 0.64\% in Dice and 1.13\% in IoU on the COVID-Xray dataset.}

\begin{figure*}[!ht]
\centering 
\includegraphics[width=0.99\linewidth]{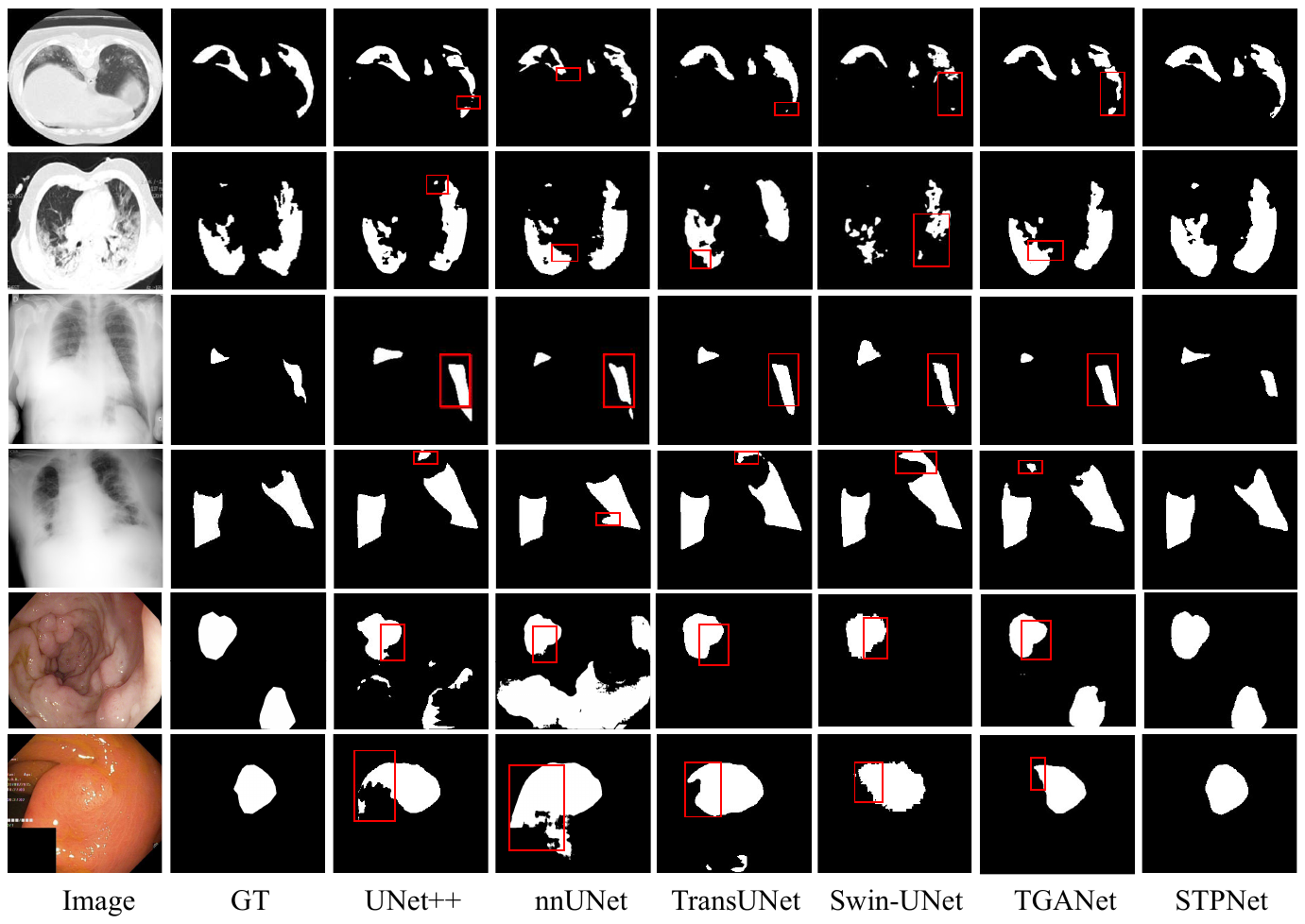}
\caption{Qualitative results on COVID-CT, COVID-Xray and Kvasir-SEG datasets. Red boxes indicate differences between \textbf{G}round \textbf{T}ruth (GT) and predictions from different baseline models.}
\label{COS}
\end{figure*}

On the Kvasir-SEG dataset, the segmentation results are shown in Table \ref{KvasirSEG}. {Our model outperformed the state-of-the-art SAM2UNet\cite{xiong2024sam2} by 5.39\% and 8.55\% in terms of Dice and IoU, respectively.} {In addition, compared to UniLSeg \cite{liu2024universal}, which requires text as input, our model achieved a Dice improvement of 7.24\%.} This demonstrates that our approach can more effectively address scale variations and perceive lesion locations. {Furthermore, to comprehensively evaluate the performance of our model, we compared Precision and Recall metrics on the Kvasir-SEG dataset. As shown in Fig. \ref{hd95}, our model's superiority in terms of Precision and Recall indicates that it effectively reduces false positives while accurately identifying target regions.}
Fig. \ref{COS} demonstrates the segmentation results on three datasets: COVID-CT, COVID-Xray, and Kvasir-SEG. The first two rows display results for COVID-CT, the middle two row for COVID-Xray, and the remaining two rows for Kvasir-SEG. The qualitative comparison results show our proposed STPNet's superiority over other state-of-the-art methods. Specifically,
for the COVID-CT dataset, other methods often exhibit incomplete segmentation in areas where the lesion scale varies, while our approach achieves relatively complete segmentation. This indicates that our method is better suited to adapt to changes in scale. As for the COVID-Xray and Kvasir-SEG datasets, other methods are prone to mis-segmentation in areas away from the lesion center, whereas our method produces segmentation boundaries that closely match the ground truth. Especially for the Kvasir-SEG dataset, which has regular lesion patterns, our method significantly outperforms other models. This suggests that our approach achieves precise lesion localization by learning positional information.
\subsection{ {Illustration of Text Retrieval Process}}

{During the text retrieval process, for each input image, we first use an image encoder to extract features and obtain the corresponding image feature vector. Meanwhile, all the texts in the text database are encoded by a text encoder and their corresponding features are stored. To retrieve the most relevant text, we compute the cosine similarity between the feature vector of the image and all the text feature vectors in four text categories (Infection text, Num text, Left Loc text, Right Loc text), and select the text with the highest similarity score in each category as the final retrieval result.
Table \ref{tab:similarity} illustrates two examples of computing similarity scores between images and all candidate texts in the COVID-Xray dataset. As shown in the first row, for this image, by computing similarity scores with the four text categories, the most similar text  retrieved from the Infection text is “Bilateral pulmonary infection”, the Num text is “Multiple infected areas”, the Left Loc text is “Upper middle lower left lung”, and the Right Loc text is “Upper middle lower right lung”.
Similarly, for the image in the second row, the most similar text  retrieved from the Infection text is “Unilateral pulmonary infection”, the Num text is “One infected area”, the Left Loc text is “No lesion in left lung”, and the Right Loc text is “Lower right lung”.
Finally, these four retrieved text features are reorganized and fed into the segmentation network, providing auxiliary semantic information to enhance segmentation performance.
}
\begin{table*}[!ht]
\centering
\caption{{Examples of computing similarity scores between images and all candidate texts in the COVID-Xray dataset. The retrieved text for each category is marked in bold.}}
\resizebox{\textwidth}{!}{%
\begin{tabular}{ccccccccc}
\hline
\textbf{Image} & \textbf{GT} & \textbf{Description} & \textbf{Infection Text (Score)} & \textbf{Num Text (Score)} & \textbf{Left Loc Text (Score)} & \textbf{Right Loc Text (Score)} \\ 
\hline

\begin{minipage}[c]{0.12\linewidth}
    \centering
    \includegraphics[width=0.9in]{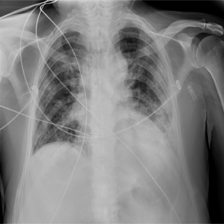}
\end{minipage} 
& \begin{minipage}[c]{0.12\linewidth}
    \centering
    \includegraphics[width=0.9in]{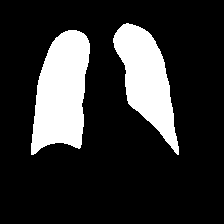}
\end{minipage}
& \makecell[c]{Bilateral pulmonary \\ infection,  two infected \\areas, all left lung and \\all right lung.
} 

& \makecell[l]{Unilateral pulmonary infection (0.0308) \\ \textbf{Bilateral pulmonary infection (0.9692)}} 

& \makecell[l]{One infected area (0.0219) \\ \textbf{Multiple infected areas (0.9781)}} 

& \makecell[l]{No lesion in left lung (0.0011) \\ Upper left lung (0.0019) \\ Middle left lung (0.0013) \\ Lower left lung (0.0107) \\ Upper lower left lung (0.0267) \\ Upper middle left lung (0.0133) \\ Middle lower left lung (0.0462) \\ \textbf{Upper middle lower left lung (0.8988)}} 

& \makecell[l]{No lesion in right lung (0.0032) \\ Upper right lung (0.0058) \\ Middle right lung (0.0044) \\ Lower right lung (0.0660) \\ Upper lower right lung (0.0563) \\ Upper middle right lung (0.0207) \\ Middle lower right lung (0.1072) \\ \textbf{Upper middle lower right lung (0.7364)}} \\

\hline

\begin{minipage}[c]{0.12\linewidth}
    \centering
    \includegraphics[width=0.9in]{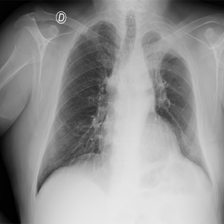}
\end{minipage}
& \begin{minipage}[c]{0.12\linewidth}
    \centering
    \includegraphics[width=0.9in]{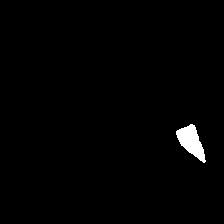}
\end{minipage}
&  \makecell[c]{Unilateral pulmonary \\ infection, one infected \\ area, lower right lung.
}

& \makecell[l]{\textbf{Unilateral pulmonary infection (0.7770
)} \\ Bilateral pulmonary infection (0.2230)}

& \makecell[l]{\textbf{One infected area (0.8004)} \\ Multiple infected areas (0.1996)} 

& \makecell[l]{\textbf{No lesion in left lung (0.5961) }\\ Upper left lung (0.0415) \\ Middle left lung (0.1665) \\ Lower left lung (0.1373) \\ Upper lower left lung (0.0093) \\ Upper middle left lung (0.0110) \\ Middle lower left lung (0.0261) \\ Upper middle lower left lung (0.0121)} 

& \makecell[l]{No lesion in right lung (0.0838) \\ Upper right lung (0.0832) \\ Middle right lung (0.3162) \\\textbf{Lower right lung (0.4222)} \\ Upper lower right lung (0.0168) \\ Upper middle right lung (0.0174) \\ Middle lower right lung (0.0439) \\ Upper middle lower right lung (0.0165)} \\

\hline
\end{tabular}%
}
\label{tab:similarity}
\end{table*}
\subsection{Ablation Studies}
In the subsection, we conduct several experiments on the COVID-CT dataset to validate the effectiveness of the proposed model, including each key component, loss function, hyperparameter settings and multi-scale text.

\textbf{Ablation studies of each key component.} 
To thoroughly evaluate the effectiveness of each component proposed in our framework, we conducted extensive ablation experiments, and the results are summarized in Table \ref{key_component}. By comparing the first and second rows, as well as the third and fourth rows of the table, we can observe that, whether it's CNNs or Transformer, the introduction of text prompts can significantly improve performance. Furthermore, the data from the second and third rows indicate that the introduction of the Transformer model can further enhance the capabilities of our framework. Lastly, incorporating SSM leads to an improvement of 0.61\% in Dice and 0.65\% in IoU. This underscores the robust advantage of SSM in capturing multiscale and spatial features.

\begin{table}[!ht]
\caption{Ablation study of key components on the COVID-CT dataset.  Text$_{CNN}$ denotes the incorporation of text information into MTBlock, while Text$_{Trans}$ represents the integration of text into UTrans.}
\normalsize
\centering
\resizebox{0.5\textwidth}{!}{%
\begin{tabular}{ccccccc}
\toprule
 \textbf{MTBlock} & \textbf{Text$_{CNN}$} & \textbf{UTrans} & \textbf{Text$_{Trans}$} & \textbf{SSM} & Dice(\%)        & IoU(\%)\\\midrule
 \checkmark & $\bigcirc$ &  $\bigcirc$ &  $\bigcirc$ & $\bigcirc$ &   74.37        &  61.54       \\
 \checkmark  & \checkmark  &  $\bigcirc$   &  $\bigcirc$ & $\bigcirc$   &   74.84        &  61.84       \\
  \checkmark  & \checkmark &  \checkmark   & $\bigcirc$  & $\bigcirc$ & 75.10         & 62.14           \\
 \checkmark  & \checkmark  &  \checkmark   & \checkmark  & $\bigcirc$  & 75.57         & 62.76           \\
 \checkmark  &  \checkmark &  \checkmark  &  \checkmark  &  \checkmark &  \textbf{76.18}         & \textbf{63.41}    
\\\bottomrule
\end{tabular}}

\label{key_component}
\end{table}

\textbf{Ablation studies of loss function.} 
As illustrated in Table \ref{ablation_loss}, we conducted a comprehensive examination of various loss functions to assess their impact on the overall performance of our model.
The results show that the gradual introduction of $\operatorname{L}_{retival}$ and $\operatorname{L}_{focal}$ consistently improves performance, highlighting the importance of accurately retrieving textual information in enhancing the accuracy of segmentation models.

\begin{table}[!ht]
\caption{Ablation studies of loss function on the COVID-CT dataset. }
\normalsize
\centering
\resizebox{0.4\textwidth}{!}{%
\begin{tabular}{ccccc}
\toprule
 \textbf{${\operatorname{L}_{img}}$} &  \textbf{${\operatorname{L}_{retrieval}}$} & \textbf{${\operatorname{L}_{focal}}$} & Dice(\%)        & IoU(\%)\\\midrule
 \checkmark &  $\bigcirc$ &  $\bigcirc$ &  75.06         & 62.21          \\
 \checkmark &  \checkmark &  $\bigcirc$ & 75.61        & 62.85 \\
 \checkmark &  \checkmark &  \checkmark & \textbf{76.18}         & \textbf{63.41}  
\\\bottomrule
\end{tabular}}

\label{ablation_loss}
\end{table}
\textbf{Ablation studies of different Hyper-Parameters.} We evaluate the effect of batch size and learning rate on the segmentation performance. For the batch size, we compared the results with values of 20, 24, and 28. For the learning rate, we set three values of 1e-4, 3e-4, and 1e-3. According to the result in Table \ref{hyper-param}, the batch size of 24 and learning rate of 3e-4 are optimal for Dice and IoU.
\begin{table}[!ht]
\caption{Ablation studies of STPNet on different Hyper-Parameters: Batch Size, Learning Rate.  The ablation experiments are conducted on the COVID-CT dataset. The Hyper-Parameters we choose are 24 for Batch Size, and 3e-4 for Learning Rate in our model.}
\small
\centering
\resizebox{0.35\textwidth}{!}{%
\begin{tabular}{cccc}
\toprule
\multicolumn{2}{l}{\textbf{Hyper-Parameters}}  & \multicolumn{1}{r}{ Dice(\%)} & \multicolumn{1}{r}{IoU(\%)} \\\midrule
\multirow{3}{*}{Batch Size} & 20  & 75.14 & 62.30 \\
                                 & 24                       & \textbf{76.18} & \textbf{63.41} \\
                                 & 28                       & 74.93 & 62.31 \\
                                 \midrule
\multirow{3}{*}{Learning Rate }  & 1e-4                     & 75.24 & 62.42 \\
                                 & {3e-4}  & \textbf{76.18} & \textbf{63.41} \\
                                 & {1e-3}  & 74.44 & 61.35\\\bottomrule
\end{tabular}}

\label{hyper-param}

\end{table}

{To further illustrate the impact of the hyper-parameters on the performance of the loss function, we performed an ablation study, the results of which are shown in Table VIII. In this study, we varied the values of $\lambda_1$, $\lambda_2$, and $\lambda_3$ to observe their effects on the model’s performance. The results indicate that when all three hyper-parameters are set to 1, the model achieves the best performance in terms of Dice score and IoU. Moreover, adjusting $\lambda_2$ has a more significant impact on the model's performance, highlighting the importance of the text retrieval process in the segmentation task. This experiment demonstrates that the loss function is quite robust and can function effectively even with small variations in these hyper-parameters.}
\vspace{-4mm}
\begin{table}[ht!]
\caption{{Ablation study on the $\lambda$ in loss function on the COVID-CT dataset.}}
\vspace{-2mm}
  \centering
  \resizebox{0.7\columnwidth}{!}{
    \begin{tabular}{ccc|cc}
    \toprule[1pt]
    $\lambda_{1}$ & $\lambda_{2}$  & $\lambda_{3}$  & Dice(\%)  & IoU(\%)  \\
   \midrule
     0.5   & 1.0  &  1.0     & 75.94 & 63.15   \\
     1.0     & 1.0 &  1.0     & \textbf{76.18}  & \textbf{63.41}  \\
     2.0     & 1.0 &  1.0     & 75.41 & 62.50   \\
    \midrule
     1.0    & 0.5  &  1.0    & 75.38  & 62.24  \\
     1.0    & 1.0 &  1.0    & \textbf{76.18}  & \textbf{63.41}  \\
     1.0    & 2.0  &  1.0    & 74.84  & 61.88  \\
    \midrule
     1.0   & 1.0 &  0.5      & 75.31 & 62.34   \\
     1.0   & 1.0 &  1.0     & \textbf{76.18}  & \textbf{63.41}  \\
     1.0   & 1.0 &  2.0      & 75.04  & 62.24  \\
   \bottomrule[1pt]
  \end{tabular}
  }
  \label{ablationl_omega}%
  \vspace{-5mm}
\end{table}


\begin{figure*}[!ht]
\centering
  \includegraphics[width=\textwidth]{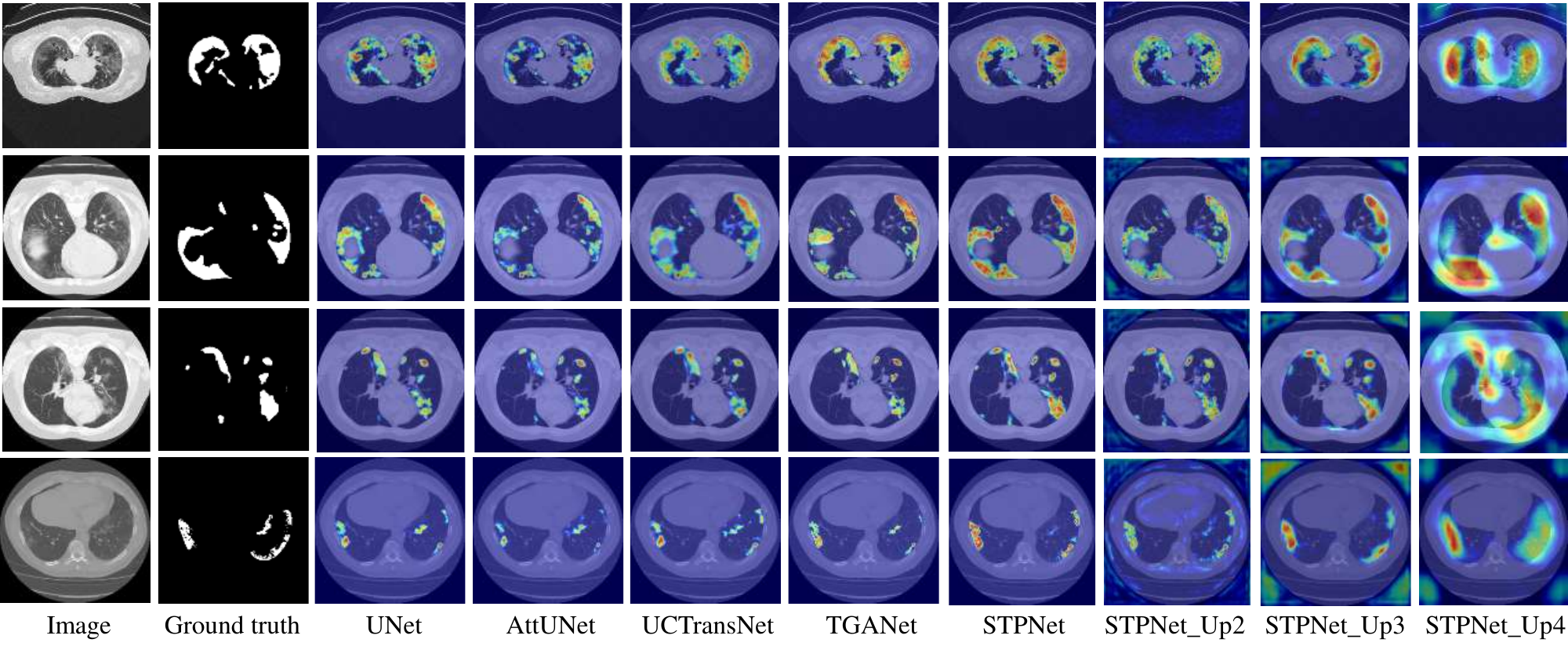}
  \vspace{-6mm}
  \caption{{Saliency maps of different approaches on the COVID-CT dataset. STPNet represents the activation map from the final layer, while STPNet$\_$Up2, STPNet$\_$Up3, and STPNet$\_$Up4 correspond to activation maps from increasingly deeper UpBlocks. The results demonstrate that integrating retrieved text information significantly reduces both mis-segmentation and missed-segmentation. Moreover, compared to TGANet, STPNet exhibits superior multi-scale feature activation. Its activation boundaries are sharper, and the highlighted regions are more precise, leading to more accurate segmentation than TGANet.}}
  \label{Inter1}
  \vspace{-4mm}
\end{figure*}

\subsection{Multi-scale Text-guided Analysis}
To investigate the impact of incorporating retrieved text types and the priority of introducing location text on model performance, we conducted relevant ablation experiments. {In cases where no text information was provided, a random vector was used as a substitute.}

Table \ref{textinsertmethod} presents the impact of introduced text types on model performance. 
Upon observing the results, it is evident that sequentially adding more detailed descriptions about the images from the shallow to the deep layers of the U-shaped network effectively enhances model performance. Specifically, when the ``infection text" representing the overall distribution of lesion locations was included, Dice improved by 0.6\%. This indicates that the overall distribution of lesion locations plays a crucial role in medical image segmentation task. When ``Num text" was added, the model performance further improved. When ``Loc text" containing information about both left and right locations was included, the model's performance also saw a significant enhancement. This improvement can be attributed to the model's gradual acquisition of more detailed text information, allowing for a more comprehensive understanding of lesion distribution within images. Therefore, systematically incorporating text information at various scales is crucial for augmenting the capabilities of deep learning models in complex data analysis and interpretation tasks.    
\vspace{-2mm}
\begin{table}[!ht]
\caption{Ablation studies of introduced text types on COVID-CT dataset. Loc text containing information about “Left Loc” and ``Right Loc".}
\vspace{-2mm}
\centering
\resizebox{0.45\textwidth}{!}{%
\begin{tabular}{ccccc}
\toprule
 \textbf{Infection text} &  \textbf{Num text} &  \textbf{Loc text} &  Dice(\%)        & IoU(\%)\\\midrule
 $\bigcirc$  & $\bigcirc$  & $\bigcirc$  &  74.64 
        & 61.67         \\ 
 \checkmark  & $\bigcirc$  & $\bigcirc$  &  75.24         & 62.31         \\ 
 \checkmark  & \checkmark  & $\bigcirc$  &  75.58         & 62.71          \\
 \checkmark & \checkmark   & \checkmark  & \textbf{76.18}         & \textbf{63.41}    
\\ \bottomrule
\end{tabular}}

\label{textinsertmethod}
\end{table}

In Fig. \ref{STPNet}, the approach presented in this paper gives priority to inserting the retrieved text information about the left location before incorporating the text information regarding the right location. To further evaluate the impact of this prioritization, we conducted experiments where we exchanged the insertion order of the two, as detailed in Table \ref{textfusethod}. Based on the experimental results, we observed that the performance in terms of the Dice coefficient obtained from both methods was comparable.
\vspace{-2mm}
\begin{table}[!ht]
\caption{The ablation of introduction priority for ``Left Loc" and ``Right Loc" on the COVID-CT dataset.}
\vspace{-1mm}
\centering
\resizebox{0.35\textwidth}{!}{%
\begin{tabular}{cccc}
\toprule
 \textbf{Left Loc} &  \textbf{Right Loc}  &  Dice(\%)        & IoU(\%)\\\midrule
 \checkmark  & $\bigcirc$   & \textbf{76.18}         & \textbf{63.41}          \\
 $\bigcirc$  & \checkmark   &75.93        & 62.97            
\\ \bottomrule
\end{tabular}}
\label{textfusethod}
\vspace{-3mm}
\end{table}
\vspace{-2mm}
\subsection{Interpretability Study}

In Fig. \ref{Inter1}, we visualized the saliency map of different approaches on the COVID-CT dataset, which vividly illustrates the multi-scale perception capability of our model. {Specifically, STPNet, STPNet$\_${Up2}, STPNet$\_${Up3}, and STPNet$\_${Up4} represent the activation maps from the decoder's UpBlocks, ranging from shallow layers to deep layers.}
Compared to other methods, our method demonstrates a heightened sensitivity to scale variations. As for both across samples with lesions of different sizes and within the same sample with lesions of varying scales, our model achieves highly accurate segmentation of the entire lesion region. Specifically, from the first row to the last row, as the lesion regions gradually decrease in size across different samples, our method consistently accurately captures the boundaries of the lesion areas.

For the first two samples, our model achieves highly precise segmentation of the entire lesion region, surpassing methods such as UNet \cite{ronneberger2015u}, AttUNet \cite{oktay2018attention}, and UCTransNet \cite{wang2022uctransnet}, which do not utilize text annotations. This underscores the advantage of incorporating text information in enhancing segmentation capabilities. Notably, when compared to the text-annotated method (TGANet \cite{huang2021gloria}), as the scale of lesions varies from small to large on the right side, the strongly activated regions of our model also adapt accordingly, whereas the changes in the strongly perceived regions of TGANet\cite{huang2021gloria} are less evident. This illustrates how our proposed multi-scale fusion approach, integrating both textual and graphical information, more effectively addresses the ``any size" issue.

{Additionally, when considering the model's layers, in the deeper layers, the focus is on the most prominent central regions of the lesion area, with attention also given to smaller lesion areas. As the network becomes shallower, the focus gradually shifts towards refining the boundary regions, progressively improving the precise localization of the lesion. This layer-by-layer optimization reflects the advantages of multi-scale feature extraction, enabling STPNet to achieve both a global understanding of the structure and precise localization of local details, thereby significantly enhancing lesion segmentation accuracy.}
\vspace{-3mm}
\subsection{{Model Computational Efficiency Analysis}}

{To offer a clear comparison of computational costs, we evaluate the inference time of STPNet against various state-of-the-art methods on the Kvasir-SEG dataset, as shown in Table \ref{KvasirSEG_infertime}.
The results show that the average inference time per image for STPNet is 0.0393s, slightly higher than some traditional methods such as AttUNet (0.0145s), but significantly lower than certain text-guided approaches like UniLSeg (0.0707s) and CMIRNet (0.1270s).  This indicates that STPNet can achieve high segmentation performance without imposing a substantial computational burden. Despite integrating text retrieval mechanisms, STPNet maintains competitive inference speed, ensuring both high segmentation accuracy and computational efficiency.}
\begin{table}[!ht]
\caption{{Average Inference Time per Image Comparison between Our Proposed Method (STPNet) and Other State-of-the-Art Methods on the Kvasir-SEG Dataset. Since our text information is retrieved by the model rather than directly added, the ``Text Input” is displayed as ``$\times$”.}}
\centering
\resizebox{0.4\textwidth}{!}{%
\begin{tabular}{cccccccc}
\toprule
{\textbf{Method}} & {\textbf{Text Input}} & Inference time(s)    \\ \midrule
AttUNet\cite{oktay2018attention}                             & $\times$   & 0.0145 \\
SAM2UNet\cite{xiong2024sam2}                           & $\times$   &
0.0330 \\
ConVIRT\cite{zhang2020contrastive}           & \checkmark & 0.0164 \\
GLoRIA\cite{huang2021gloria}                 & \checkmark & 0.0159 \\
CLIP\cite{radford2021learning}             & \checkmark &  0.0173\\
ViLT\cite{28kim2021vilt}                   & \checkmark & 0.0405 \\
LAVT\cite{29yang2021lavt}                 & \checkmark & 0.0447\\
UniLSeg
\cite{liu2024universal}   & \checkmark & 0.0707 \\

CMIRNet
\cite{xu2024cmirnet}   & \checkmark & 0.1270 \\
\textbf{STPNet (ours)}              & $\times$   &0.0393 \\\bottomrule
\end{tabular}
}

\label{KvasirSEG_infertime}
\vspace{-4mm}
\end{table}


\section{Conclusion}
\label{sec:conclusion}
Lesions can manifest in different locations and sizes, making it crucial to consider both factors simultaneously for enhancing segmentation performance. Our proposed STPNet achieves this by retrieving text features containing location information and recombining them into multi-scale semantic text features. STPNet consists of a text retrieval network and a segmentation network. Firstly, the text retrieval network retrieves text features based on image features, which are then recombined and input into the MTBlock and UTrans of the segmentation network to facilitate interaction between different modalities. Comprehensive experiments have demonstrated that our STPNet can achieve superior segmentation performance compared to current state-of-the-art methods on two pneumonia datasets, including two modalities: CT and X-ray, as well as one polyp dataset. In addition, STPNet eliminates the need for additional textual information during inference in multimodal models, which greatly enhances the model's adaptability and practicality.

{It should also be noted that our model's effectiveness may be influenced by the quality of the specialized medical text repository. The performance disparities between the Kvasir-SEG dataset (98.19\% Dice) and pneumonia datasets (76.18\% and 80.63\% Dice) suggest that STPNet performs better on datasets with more regular lesion patterns, indicating the potential improvement in handling highly irregular morphologies with ambiguous boundaries. Additionally, from a computational standpoint, STPNet introduces additional overhead compared to purely image-based approaches, which may pose challenges for deployment in resource-limited clinical environments. Future work should focus on incorporating uncertainty estimation to enhance clinical trust, developing adaptable text repositories that evolve with new knowledge, exploring parameter-efficient transformer variants to reduce computational costs, and extending the framework to incorporate disease classification and severity assessment for more comprehensive diagnostic support.}

\bibliographystyle{IEEEtran}
\bibliography{main}

\begin{thebibliography}{10}
\providecommand{\url}[1]{#1}
\csname url@samestyle\endcsname
\providecommand{\newblock}{\relax}
\providecommand{\bibinfo}[2]{#2}
\providecommand{\BIBentrySTDinterwordspacing}{\spaceskip=0pt\relax}
\providecommand{\BIBentryALTinterwordstretchfactor}{4}
\providecommand{\BIBentryALTinterwordspacing}{\spaceskip=\fontdimen2\font plus
\BIBentryALTinterwordstretchfactor\fontdimen3\font minus \fontdimen4\font\relax}
\providecommand{\BIBforeignlanguage}[2]{{%
\expandafter\ifx\csname l@#1\endcsname\relax
\typeout{** WARNING: IEEEtran.bst: No hyphenation pattern has been}%
\typeout{** loaded for the language `#1'. Using the pattern for}%
\typeout{** the default language instead.}%
\else
\language=\csname l@#1\endcsname
\fi
#2}}
\providecommand{\BIBdecl}{\relax}
\BIBdecl

\bibitem{fan2020inf}
D.-P. Fan, T.~Zhou, G.-P. Ji, Y.~Zhou, G.~Chen, H.~Fu, J.~Shen, and L.~Shao, ``Inf-net: Automatic covid-19 lung infection segmentation from ct images,'' \emph{IEEE Transactions on Medical Imaging}, vol.~39, no.~8, pp. 2626--2637, 2020.

\bibitem{zhao2021scoat}
S.~Zhao, Z.~Li, Y.~Chen, W.~Zhao, X.~Xie, J.~Liu, D.~Zhao, and Y.~Li, ``Scoat-net: A novel network for segmenting covid-19 lung opacification from ct images,'' \emph{Pattern Recognition}, vol. 119, p. 108109, 2021.

\bibitem{huang2021graph}
H.~Huang, L.~Lin, Y.~Zhang, Y.~Xu \emph{et~al.}, ``Graph-bas3net: Boundary-aware semi-supervised segmentation network with bilateral graph convolution,'' in \emph{Proceedings of the IEEE/CVF International Conference on Computer Vision}, 2021, pp. 7386--7395.

\bibitem{li2022tfcns}
Z.~Li, D.~Li, C.~Xu \emph{et~al.}, ``Tfcns: A cnn-transformer hybrid network for medical image segmentation,'' in \emph{International Conference on Artificial Neural Networks}.\hskip 1em plus 0.5em minus 0.4em\relax Springer, 2022, pp. 781--792.

\bibitem{xu2022mrdff}
F.~Xu, L.~Lin, Z.~Li, Q.~Hong, K.~Liu, Q.~Wu, Q.~Li, Y.~Zheng, and J.~Tian, ``Mrdff: A deep forest based framework for ct whole heart segmentation,'' \emph{Methods}, vol. 208, pp. 48--58, 2022.

\bibitem{yan2021covid}
Q.~Yan, B.~Wang, D.~Gong \emph{et~al.}, ``Covid-19 chest ct image segmentation network by multi-scale fusion and enhancement operations,'' \emph{IEEE transactions on big data}, vol.~7, no.~1, pp. 13--24, 2021.

\bibitem{zheng2020msd}
B.~Zheng, Y.~Liu, Y.~Zhu, F.~Yu, T.~Jiang, D.~Yang, and T.~Xu, ``Msd-net: Multi-scale discriminative network for covid-19 lung infection segmentation on ct,'' \emph{Ieee Access}, vol.~8, pp. 185\,786--185\,795, 2020.

\bibitem{pei2021mps}
H.-Y. Pei, D.~Yang, G.-R. Liu, and T.~Lu, ``Mps-net: Multi-point supervised network for ct image segmentation of covid-19,'' \emph{Ieee Access}, vol.~9, pp. 47\,144--47\,153, 2021.

\bibitem{10172039}
Z.~Li, Y.~Li, Q.~Li, P.~Wang, D.~Guo, L.~Lu, D.~Jin, Y.~Zhang, and Q.~Hong, ``Lvit: Language meets vision transformer in medical image segmentation,'' \emph{IEEE Transactions on Medical Imaging}, vol.~43, no.~1, pp. 96--107, 2024.

\bibitem{tomar2022tganet}
N.~K. Tomar, D.~Jha, U.~Bagci, and S.~Ali, ``Tganet: text-guided attention for improved polyp segmentation,'' in \emph{Medical Image Computing and Computer Assisted Intervention--MICCAI 2022: 25th International Conference, Singapore, September 18--22, 2022, Proceedings, Part III}.\hskip 1em plus 0.5em minus 0.4em\relax Springer, 2022, pp. 151--160.

\bibitem{huemann2024contextual}
Z.~Huemann, X.~Tie, J.~Hu, and T.~J. Bradshaw, ``Contextual net: a multimodal vision-language model for segmentation of pneumothorax,'' \emph{Journal of Imaging Informatics in Medicine}, pp. 1--12, 2024.

\bibitem{zhong2023ariadne}
Y.~Zhong, M.~Xu, K.~Liang, K.~Chen, and M.~Wu, ``Ariadne’s thread: Using text prompts to improve segmentation of infected areas from chest x-ray images,'' in \emph{International Conference on Medical Image Computing and Computer-Assisted Intervention}.\hskip 1em plus 0.5em minus 0.4em\relax Springer, 2023, pp. 724--733.

\bibitem{ronneberger2015u}
O.~Ronneberger, P.~Fischer, and T.~Brox, ``U-net: Convolutional networks for biomedical image segmentation,'' in \emph{International Conference on Medical image computing and computer-assisted intervention}.\hskip 1em plus 0.5em minus 0.4em\relax Springer, 2015, pp. 234--241.

\bibitem{valanarasu2022unext}
J.~M.~J. Valanarasu and V.~M. Patel, ``Unext: Mlp-based rapid medical image segmentation network,'' in \emph{International Conference on Medical Image Computing and Computer-Assisted Intervention}.\hskip 1em plus 0.5em minus 0.4em\relax Springer, 2022, pp. 23--33.

\bibitem{han2022convunext}
Z.~Han, M.~Jian, and G.-G. Wang, ``Convunext: An efficient convolution neural network for medical image segmentation,'' \emph{Knowledge-Based Systems}, vol. 253, p. 109512, 2022.

\bibitem{zhou2018unet++}
Z.~Zhou, M.~M.~R. Siddiquee, N.~Tajbakhsh, and J.~Liang, ``Unet++: Redesigning skip connections to exploit multiscale features in image segmentation,'' \emph{IEEE transactions on medical imaging}, vol.~39, no.~6, pp. 1856--1867, 2019.

\bibitem{oktay2018attention}
O.~Oktay, J.~Schlemper, L.~L. Folgoc, M.~Lee, M.~Heinrich, K.~Misawa, K.~Mori, S.~McDonagh, N.~Y. Hammerla, B.~Kainz \emph{et~al.}, ``Attention u-net: Learning where to look for the pancreas,'' \emph{arXiv preprint arXiv:1804.03999}, 2018.

\bibitem{isensee2021nnu}
F.~Isensee, P.~F. Jaeger, S.~A. Kohl, J.~Petersen, and K.~H. Maier-Hein, ``nnu-net: a self-configuring method for deep learning-based biomedical image segmentation,'' \emph{Nature methods}, vol.~18, no.~2, pp. 203--211, 2021.

\bibitem{cao2021swin}
H.~Cao, Y.~Wang, J.~Chen, D.~Jiang, X.~Zhang, Q.~Tian, and M.~Wang, ``Swin-unet: Unet-like pure transformer for medical image segmentation,'' in \emph{Computer Vision--ECCV 2022 Workshops: Tel Aviv, Israel, October 23--27, 2022, Proceedings, Part III}.\hskip 1em plus 0.5em minus 0.4em\relax Springer, 2022, pp. 205--218.

\bibitem{xiong2024sam2}
X.~Xiong, Z.~Wu, S.~Tan, W.~Li, F.~Tang, Y.~Chen, S.~Li, J.~Ma, and G.~Li, ``Sam2-unet: Segment anything 2 makes strong encoder for natural and medical image segmentation,'' \emph{arXiv preprint arXiv:2408.08870}, 2024.

\bibitem{ravi2024sam}
N.~Ravi, V.~Gabeur, Y.-T. Hu, R.~Hu, C.~Ryali, T.~Ma, H.~Khedr, R.~R{\"a}dle, C.~Rolland, L.~Gustafson \emph{et~al.}, ``Sam 2: Segment anything in images and videos,'' \emph{arXiv preprint arXiv:2408.00714}, 2024.

\bibitem{you2022class}
C.~You, R.~Zhao, F.~Liu, S.~Dong, S.~Chinchali, U.~Topcu, L.~Staib, and J.~Duncan, ``Class-aware adversarial transformers for medical image segmentation,'' \emph{Advances in Neural Information Processing Systems}, vol.~35, pp. 29\,582--29\,596, 2022.

\bibitem{wu2023d}
Y.~Wu, K.~Liao, J.~Chen, J.~Wang, D.~Z. Chen, H.~Gao, and J.~Wu, ``D-former: A u-shaped dilated transformer for 3d medical image segmentation,'' \emph{Neural Computing and Applications}, vol.~35, no.~2, pp. 1931--1944, 2023.

\bibitem{hatamizadeh2022unetr}
A.~Hatamizadeh, Y.~Tang, V.~Nath, D.~Yang, A.~Myronenko, B.~Landman, H.~R. Roth, and D.~Xu, ``Unetr: Transformers for 3d medical image segmentation,'' in \emph{Proceedings of the IEEE/CVF winter conference on applications of computer vision}, 2022, pp. 574--584.

\bibitem{chen2021transunet}
J.~Chen, Y.~Lu, Q.~Yu, X.~Luo, E.~Adeli, Y.~Wang, L.~Lu, A.~L. Yuille, and Y.~Zhou, ``Transunet: Transformers make strong encoders for medical image segmentation,'' \emph{arXiv preprint arXiv:2102.04306}, 2021.

\bibitem{bakkouri2023mlca2f}
I.~Bakkouri and K.~Afdel, ``Mlca2f: Multi-level context attentional feature fusion for covid-19 lesion segmentation from ct scans,'' \emph{Signal, Image and Video Processing}, vol.~17, no.~4, pp. 1181--1188, 2023.

\bibitem{LIU2023107493}
\BIBentryALTinterwordspacing
Y.~Liu, Y.~Zhu, Y.~Xin, Y.~Zhang, D.~Yang, and T.~Xu, ``Mestrans: Multi-scale embedding spatial transformer for medical image segmentation,'' \emph{Computer Methods and Programs in Biomedicine}, vol. 233, p. 107493, 2023. [Online]. Available: \url{https://www.sciencedirect.com/science/article/pii/S0169260723001591}
\BIBentrySTDinterwordspacing

\bibitem{ding2022vlt}
H.~Ding, C.~Liu, S.~Wang, and X.~Jiang, ``Vlt: Vision-language transformer and query generation for referring segmentation,'' \emph{IEEE Transactions on Pattern Analysis and Machine Intelligence}, 2022.

\bibitem{yang2022lavt}
Z.~Yang, J.~Wang, Y.~Tang, K.~Chen, H.~Zhao, and P.~H. Torr, ``Lavt: Language-aware vision transformer for referring image segmentation,'' in \emph{Proceedings of the IEEE/CVF Conference on Computer Vision and Pattern Recognition}, 2022, pp. 18\,155--18\,165.

\bibitem{liu2024universal}
Y.~Liu, C.~Zhang, Y.~Wang, J.~Wang, Y.~Yang, and Y.~Tang, ``Universal segmentation at arbitrary granularity with language instruction,'' in \emph{Proceedings of the IEEE/CVF Conference on Computer Vision and Pattern Recognition}, 2024, pp. 3459--3469.

\bibitem{xu2024cmirnet}
M.~Xu, T.~Xiao, Y.~Liu, H.~Tang, Y.~Hu, and L.~Nie, ``Cmirnet: Cross-modal interactive reasoning network for referring image segmentation,'' \emph{IEEE Transactions on Circuits and Systems for Video Technology}, 2025.

\bibitem{shan2023coarse}
D.~Shan, Z.~Li, W.~Chen, Q.~Li, J.~Tian, and Q.~Hong, ``Coarse-to-fine covid-19 segmentation via vision-language alignment,'' in \emph{ICASSP 2023-2023 IEEE International Conference on Acoustics, Speech and Signal Processing (ICASSP)}.\hskip 1em plus 0.5em minus 0.4em\relax IEEE, 2023, pp. 1--5.

\bibitem{poudel2023exploring}
K.~Poudel, M.~Dhakal, P.~Bhandari, R.~Adhikari, S.~Thapaliya, and B.~Khanal, ``Exploring transfer learning in medical image segmentation using vision-language models,'' \emph{arXiv preprint arXiv:2308.07706}, 2023.

\bibitem{radford2021learning}
A.~Radford, J.~W. Kim, C.~Hallacy \emph{et~al.}, ``Learning transferable visual models from natural language supervision,'' in \emph{International Conference on Machine Learning}.\hskip 1em plus 0.5em minus 0.4em\relax PMLR, 2021, pp. 8748--8763.

\bibitem{huang2021gloria}
S.-C. Huang, L.~Shen, M.~P. Lungren, and S.~Yeung, ``Gloria: A multimodal global-local representation learning framework for label-efficient medical image recognition,'' in \emph{Proceedings of the IEEE/CVF International Conference on Computer Vision}, 2021, pp. 3942--3951.

\bibitem{zhang2020contrastive}
Y.~Zhang, H.~Jiang, Y.~Miura, C.~D. Manning, and C.~P. Langlotz, ``Contrastive learning of medical visual representations from paired images and text,'' pp. 2--25, 2022.

\bibitem{he2016deep}
K.~He, X.~Zhang, S.~Ren, and J.~Sun, ``Deep residual learning for image recognition,'' in \emph{Proceedings of the IEEE conference on computer vision and pattern recognition}, 2016, pp. 770--778.

\bibitem{alsentzer2019publicly}
E.~Alsentzer, J.~R. Murphy, W.~Boag, W.-H. Weng, D.~Jin, T.~Naumann, and M.~McDermott, ``Publicly available clinical bert embeddings,'' \emph{arXiv preprint arXiv:1904.03323}, 2019.

\bibitem{hu2023animate}
L.~Hu, X.~Gao, P.~Zhang, K.~Sun, B.~Zhang, and L.~Bo, ``Animate anyone: Consistent and controllable image-to-video synthesis for character animation,'' \emph{arXiv preprint arXiv:2311.17117}, 2023.

\bibitem{ma2020towards}
J.~Ma, Y.~Wang, X.~An, C.~Ge, Z.~Yu, J.~Chen, Q.~Zhu, G.~Dong, J.~He, Z.~He \emph{et~al.}, ``Towards efficient covid-19 ct annotation: A benchmark for lung and infection segmentation,'' \emph{arXiv preprint arXiv:2004.12537}, vol.~9, 2020.

\bibitem{morozov2020mosmeddata}
S.~P. Morozov, A.~Andreychenko, N.~Pavlov, A.~Vladzymyrskyy, N.~Ledikhova, V.~Gombolevskiy, I.~A. Blokhin, P.~Gelezhe, A.~Gonchar, and V.~Y. Chernina, ``Mosmeddata: Chest ct scans with covid-19 related findings dataset,'' \emph{arXiv preprint arXiv:2005.06465}, 2020.

\bibitem{degerli2021covid}
A.~Degerli, M.~Ahishali, M.~Yamac, S.~Kiranyaz, M.~E. Chowdhury, K.~Hameed, T.~Hamid, R.~Mazhar, and M.~Gabbouj, ``Covid-19 infection map generation and detection from chest x-ray images,'' \emph{Health information science and systems}, vol.~9, no.~1, p.~15, 2021.

\bibitem{jha2020kvasir}
D.~Jha, P.~H. Smedsrud, M.~A. Riegler, P.~Halvorsen, T.~de~Lange, D.~Johansen, and H.~D. Johansen, ``Kvasir-seg: A segmented polyp dataset,'' in \emph{International Conference on Multimedia Modeling}.\hskip 1em plus 0.5em minus 0.4em\relax Springer, 2020, pp. 451--462.

\bibitem{wang2022uctransnet}
H.~Wang, P.~Cao, J.~Wang, and O.~R. Zaiane, ``Uctransnet: rethinking the skip connections in u-net from a channel-wise perspective with transformer,'' in \emph{Proceedings of the AAAI Conference on Artificial Intelligence}, vol.~36, no.~3, 2022, pp. 2441--2449.

\bibitem{28kim2021vilt}
W.~Kim, B.~Son, and I.~Kim, ``Vilt: Vision-and-language transformer without convolution or region supervision,'' in \emph{International Conference on Machine Learning}.\hskip 1em plus 0.5em minus 0.4em\relax PMLR, 2021, pp. 5583--5594.

\bibitem{29yang2021lavt}
Z.~Yang, J.~Wang, Y.~Tang, K.~Chen, H.~Zhao, and P.~H. Torr, ``Lavt: Language-aware vision transformer for referring image segmentation,'' \emph{arXiv preprint arXiv:2112.02244}, 2021.

\end{thebibliography}

\vfill

\end{document}